\documentclass[aps,prb,twocolumn,longbibliography,amsmath,amssymb,superscriptaddress, bibnotes]{revtex4-2}

\usepackage{graphicx}
\usepackage{dcolumn}
\usepackage{bm}
\usepackage{color}
\usepackage{ulem}

\begin{document}

\title{Intrinsic low-temperature magnetic properties\\ on the ultra-clean UTe$_2$ with $T_{\rm c}$ = 2.1 K revealed by $^{125}$Te NMR}

\author{Hiroki~Matsumura}
\email{matsumura.hiroki.75r@st.kyoto-u.ac.jp}
\author{Shunsaku~Kitagawa}
\email{kitagawa.shunsaku.8u@kyoto-u.ac.jp}
\author{Shiki~Ogata}
\author{Riku~Matsubayashi}
\author{Hiroki~Fujibayashi}
\affiliation{Department of Physics, Graduate School of Science, Kyoto University, Kyoto 606-8502, Japan}

\author{Katsuki~Kinjo}
\affiliation{Department of Physics, Graduate School of Science, Kyoto University, Kyoto 606-8502, Japan}
\affiliation{Institute of Multidisciplinary Research for Advanced Materials, Tohoku University, Sendai, Miyagi 980-8577, Japan}

\author{Kenji~Ishida}
\affiliation{Department of Physics, Graduate School of Science, Kyoto University, Kyoto 606-8502, Japan}

\author{Yo~Tokunaga}
\author{Hironori~Sakai} 
\author{Shinsaku~Kambe}
\affiliation{Advanced Science Research Center, Japan Atomic Energy Agency, Tokai, Ibaraki 319-1195, Japan}

\author{Ai~Nakamura}
\author{Yusei~Shimizu}
\author{Yoshiya~Homma}
\author{Dexin~Li}
\affiliation{Institute for Materials Research, Tohoku University, Oarai, Ibaraki 311-1313, Japan}

\author{Fuminori~Honda}
\affiliation{Institute for Materials Research, Tohoku University, Oarai, Ibaraki 311-1313, Japan}
\affiliation{Central Institute of Radioisotope Science and Safety, Kyushu University, Fukuoka 819-0395, Japan}

\author{Atsushi~Miyake}
\affiliation{Institute for Materials Research, Tohoku University, Oarai, Ibaraki 311-1313, Japan}

\author{Dai~Aoki}
\affiliation{Institute for Materials Research, Tohoku University, Oarai, Ibaraki 311-1313, Japan}
\affiliation{Universit\'e Grenoble Alpes, CEA, IRIG, PHELIQS, F-38000 Grenoble, France}

\date{\today}

\begin{abstract}
To investigate the intrinsic magnetic properties of UTe$_2$, we performed $^{125}$Te-NMR measurements on the ultra-clean single-crystalline UTe$_2$ with superconducting transition temperature $T_{\rm c}$ = 2.1~K and compared the results with those of the $T_{\rm c}$ = 1.6~K sample.
The broadening of the linewidth of the NMR spectrum in the $a$-axis magnetic field and the low-temperature magnetic fluctuations observed in the 1.6~K sample are suppressed in the ultra-clean sample, indicating that such magnetic properties originate from a tiny amount of U deficiency.
The present results suggest that the magnetic properties in UTe$_2$ are sensitive to the U deficiency.
We also observed a peculiar angular dependence of the NMR quantities due to large magnetic anisotropy with the $a$-axis as the magnetic easy axis. 
\end{abstract}

\maketitle

\section{Introduction}
In conventional superconductors, the electron-phonon interactions are the origin of an attractive force for the Cooper pairs\cite{J.Bardeen_PR_1957}.
On the other hand, in unconventional superconductors, the attractive force is not mediated by phonon, but by other interactions, such as magnetic interactions\cite{T.Moriya_AP_2000,T.Moriya_RPP_2003,P.Monthoux_PRL_1991,C.Pfleiderer_RMP_2009,Y.Nakai_PRB_2013,S.Kitagawa_PRB_2019,V.P.Mineev_PU_2017,M.Manago_PRB_2019,K.Ishida_PRB_2021,S.Kitagawa_JPSJ_2022,S.Ogawa_JPSJ_2023,H.Sato_JPSJ_2023,K.Omasa_JPSJ_2024}.
This also seems to be the case in the recently discovered superconductor UTe$_2$.
UTe$_2$ is a uranium-based compound with a superconducting (SC) transition temperature of $T_{\rm c}$ = 1.6 - 2.1 K\cite{S.Ran_Science_2019,D.Aoki_JPSJ_2019_2,D.Aoki_JPCondMatt_2022}.
Although it does not exhibit a ferromagnetic transition, it shows many similarities to ferromagnetic superconductors\cite{S.S.Saxena_Nature_2000,D.Aoki_nature_2001,N.T.Huy_PRL_2007,D.Aoki_JPSJ_2019}, such as Ising magnetic anisotropy and a peculiar SC phase diagram\cite{S.Ran_Science_2019,G.Knebel_JPSJ_2019,G.Knebel_JPSJ_2020,A.Rosuel_PRX_2023,D.Braithwaite_CommunPhys_2019,D.Aoki_JPSJ_2020,S.Ran_NatPhys_2019,W.Knafo_CommPhys_2021,H.Sakai_PRL_2023,K.Kinjo_PRB_2023,K.Kinjo_SciAdv_2023,M.Ajeesh_JPSJ_2024,D.Aoki_JPSJ_2024b}.
Thus, when it was first discovered, the realization of a spin-triplet SC state mediated by ferromagnetic fluctuations was pointed out\cite{S.Ran_Science_2019}.
In fact, strong ferromagnetic fluctuations were suggested from various measurements\cite{S.Ran_Science_2019,Y.Tokunaga_JPSJ_2019,S.Sundar_PRB_2019,Y.Tokunaga_PRL_2023}.
On the other hand, inelastic neutron scattering experiments suggested the existence of antiferromagnetic fluctuations with $Q$ = (0, 0.57, 0) at ambient pressure\cite{C.Duan_PRL_2020,W.Knafo_PRB_2021}.
In addition, static antiferromagnetic order with $Q$ = (0.07, 0.67, 0) was found under pressure\cite{W.Knafo_arXiv_2023}, implying that the magnetic properties of the normal state in UTe$_2$ are not simple.
In addition, NMR measurements in the early-stage samples with $T_{\rm c} \sim 1.6$ K (1.6~K sample) suggested that antiferromagnetic-type fluctuations develop at low temperatures due to a tiny amount of U deficiency\cite{Y.Tokunaga_JPSJ_2022,H.Fujibayashi_JPSJ_2023}.
$\mu$SR measurements also detected the spin freezing in UTe$_2$ samples with the residual electronic term in the specific heat\cite{S.Sundar_PRB_2019,S.Sundar_CommPhys_2023}.
Therefore, high-quality samples are necessary to investigate the intrinsic magnetic properties of UTe$_2$.

\begin{figure}[!tb]
\includegraphics[width=8.5cm,clip]{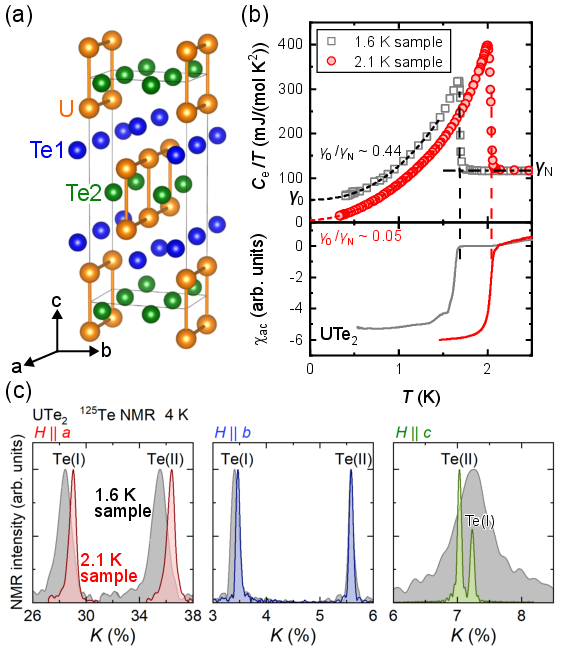}
\caption{
(a) Crystal structure of UTe$_2$ drawn by VESTA\cite{K.Momma_JAC_2011}.
(b) Temperature dependence of the electronic term of the speciﬁc heat divided by temperature $C_e/T$ and the ac susceptibility measured with variation in resonance frequency of NMR tank circuits in the 1.6 and 2.1 K samples.
The vertical and horizontal broken lines indicate $T_{\rm c}$ and the normal-state Sommerfeld coefficient $\gamma_{\rm N}$, respectively.
Residual-specific heat values for each sample, scaled by the normal state value, $\gamma_{0}/\gamma_{\rm N}$ were estimated by extrapolation of the full-gap behavior\cite{H.Matsumura_JPSJ_2023}.
(c) $^{125}$Te-NMR spectrum as a function of $K$ measured in the 1.6 and 2.1 K samples when $H$ is applied to three crystalline axes ($a, b$, and $c$).
In $H \parallel c$, the 1.6 K sample shows the broad spectrum due to overlapping Te(I) and Te(II)-NMR peaks, but the two peaks were separated in the 2.1 K sample.
All spectra were measured at 4 K.
}
\label{Fig.1}
\end{figure}

Recently, an ultra-clean single-crystal sample (2.1~K sample) with a large increase in $T_{\rm c}$ to 2.1~K and almost zero residual electronic coefficient at 0~K is available due to an improvement in the synthesis method\cite{H.Sakai_PRM_2022}, as shown in Fig.~\ref{Fig.1}(b).
A low-temperature upturn in magnetic susceptibility for $H\parallel a$ observed in the 1.6~K sample\cite{S.Ran_Science_2019,D.Li_JPSJ_2021} is absent in the 2.1 K sample\cite{D.Aoki_JPSJ_2024}. 
In addition to the magnetic properties, the relationship between $T_{\rm c}$ and the residual electronic term in the specific heat indicates that the 2.1~K sample is a disorder-free sample with intrinsic $T_{\rm c}$ in UTe$_2$\cite{D.Aoki_JPCondMatt_2022,D.Aoki_JPSJ_2024}.

In this paper, we perform NMR measurements on this ultra-clean single-crystal sample and compare the results with those of the 1.6~K sample to investigate the intrinsic magnetic properties of UTe$_2$.
Our NMR measurements reveal that the low-temperature magnetic anomalies observed in the 1.6~K sample are not observed in the 2.1~K samples.

\section{Experimental}
$^{125}$Te-enriched ultra-clean samples were prepared with the newly developed molten salt flux method\cite{H.Sakai_PRM_2022}.
Natural U and 99.9\% $^{125}$Te-enriched metals were used as starting materials for the present sample.
The ferromagnetic U$_7$Te$_{12}$\cite{P.Opletal_JPSJ_2023} is absent in our samples.
The single crystals exhibit an SC transition at $T_{\rm c} = 2.1$~K, which was determined with the specific heat and ac susceptibility measurements as shown in Fig.~\ref{Fig.1}(b)\cite{D.Aoki_JPSJ_2022b,H.Matsumura_JPSJ_2023} .
This is the highest $T_{\rm c}$ reported in UTe$_2$\cite{D.Aoki_JPCondMatt_2022,D.Aoki_JPSJ_2024}.
This sample shows a quite small residual electronic term well below $T_{\rm c}$, indicating that the increase in $T_{\rm c}$ is ascribed to the reduction of the disorder due to the U deficiency\cite{Y.Haga_JPCM_2022,H.Sakai_PRM_2022}.
For comparison, we also performed NMR measurements in the early-stage 1.6~K sample grown by the chemical vapor transport method\cite{D.Aoki_JPSJ_2019_2}.
The 1.6 K sample is identical to that used in our previous study, and some of the data were already published\cite{H.Fujibayashi_JPSJ_2022,S.Kitagawa_JPSJ_2024}.
The $^{125}$Te (nuclear spin $I = 1/2$, gyromagnetic ratio $^{125}\gamma/2\pi = 13.454$~MHz/T)-NMR measurements on the 2.1 K sample were performed on two same-batch single crystals of size 2 $\times$ 1.2 $\times$ 1 mm$^3$ for the $ab$ and $ac$ plane rotations and 3 $\times$ 1 $\times$ 0.5 mm$^3$ for the $bc$ plane rotation.
We reported that two $^{125}$Te-NMR signals were observed in UTe$_2$ due to the presence of the two inequivalent crystallographic Te sites, as seen in Fig.~\ref{Fig.1}(a).
Note that the peak assignment [Te(I) and Te(II)] does not correspond to the crystallographic site (Te1 and Te2)\cite{H.Fujibayashi_JPSJ_2023}.
The NMR spectra as a function of frequency were obtained using the Fourier transform of a spin–echo signal observed after a radio-frequency pulse sequence at a fixed magnetic field.
The NMR spectra are presented with the horizontal axis representing $K = (f - f_0)/f_0$.
Here, $f_0 = (^{125}\gamma/2\pi)\mu_0H$ is the reference frequency.
The magnetic field was calibrated using a $^{65}$Cu ($^{65}\gamma/2\pi = 12.089$~MHz/T)-NMR signal with the Knight shift $K_{\rm Cu} = 0.2385$\% from the NMR coil\cite{MetallicShift}.
The sample was rotated in the $ab$, $bc$, and $ac$ planes to precisely apply the magnetic field along each axis using a split-pair magnet with a single-axis rotator.
The Knight shift and full width at half maximum (FWHM) were determined by the peak position and width at the half-maximum value of the NMR spectrum, respectively.
In $H \parallel c$, the NMR signals of the two sites are quite close.
Therefore, the FWHM was determined by fitting with a double Gaussian.
A nuclear spin-lattice relaxation rate $1/T_1$ was evaluated by fitting the relaxation curve of the nuclear magnetization after the saturation to a theoretical function for the nuclear spin $I = 1/2$, which is a single exponential function.

\section{Results and Discussion}
\subsection{Knighit shift and anomalous broadening of NMR linewidth}
\begin{figure}[!tb]
\centering
\includegraphics[width=\linewidth,clip]{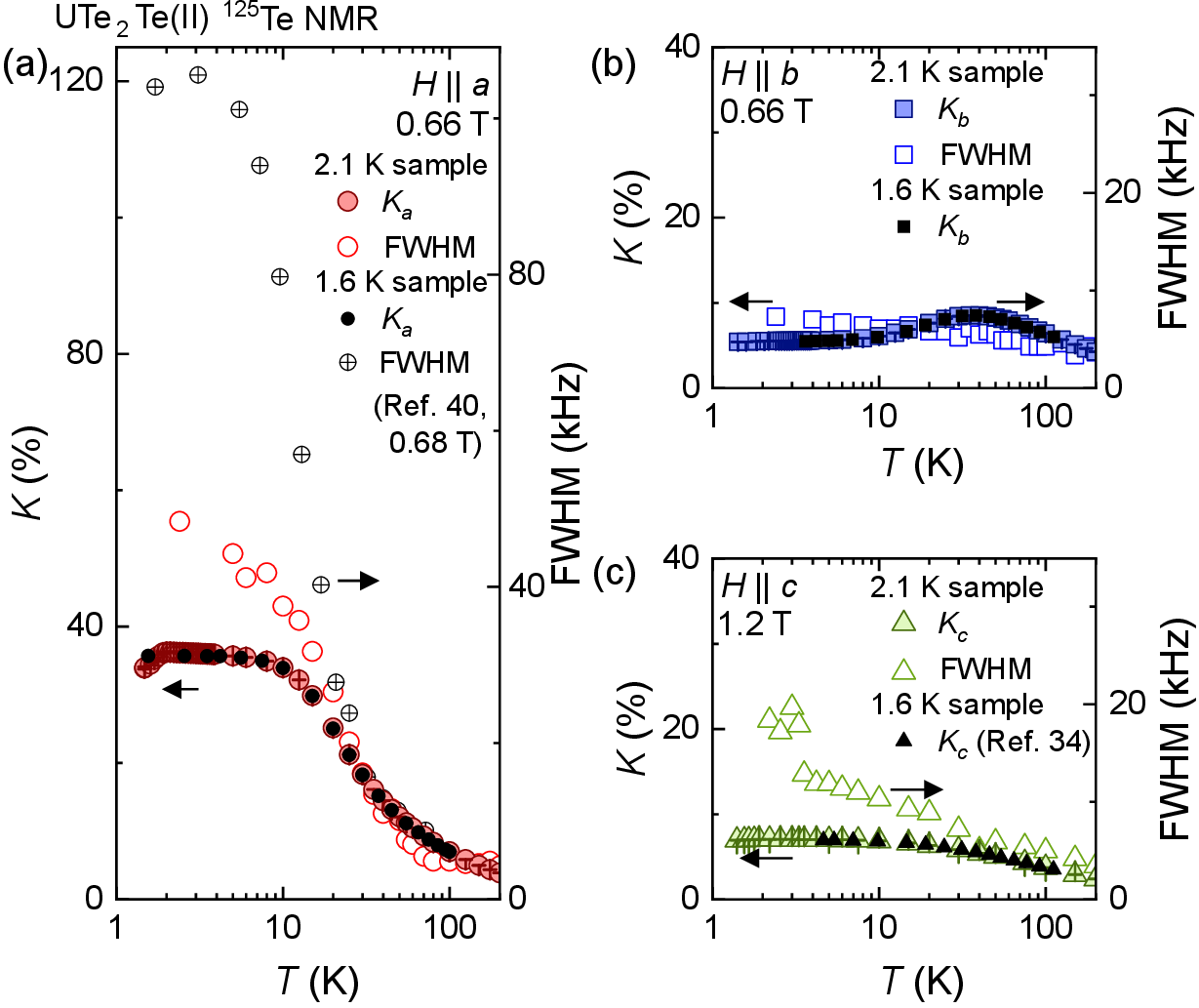}
\caption{
Temperature dependence of Knight shift $K$ and full width at half maximum (FWHM) at the Te(II) site for (a)$H \parallel a$, (b)$H \parallel b$, and (c)$H \parallel c$.
We also plot the data in the 1.6~K sample for comparison.
}
\label{Fig.2}
\end{figure}

As shown in Fig.~\ref{Fig.1} (c), the NMR spectra for the 2.1~K sample are sharper in all crystallographic axes ($a, b$, and $c$) than those for the 1.6 K sample\cite{G.Nakamine_PRB_2021,H.Fujibayashi_JPSJ_2022}, indicating the improvement of the sample quality.
In particular, note that the sharpening of the NMR spectrum along the $c$ axis separates between the Te(I) and Te(II) peaks.
In contrast to linewidth, as shown in Figs.~\ref{Fig.2}(a)-(c), the Knight shifts for both the 1.6 K and 2.1 K samples exhibit almost the same behavior, implying no significant change in static magnetic properties.
Slight differences in peak positions between the samples [Fig.~\ref{Fig.1} (c)] may originate from misalignment of field direction.

\begin{figure}[!tb]
\centering
\includegraphics[width=8.5cm,clip]{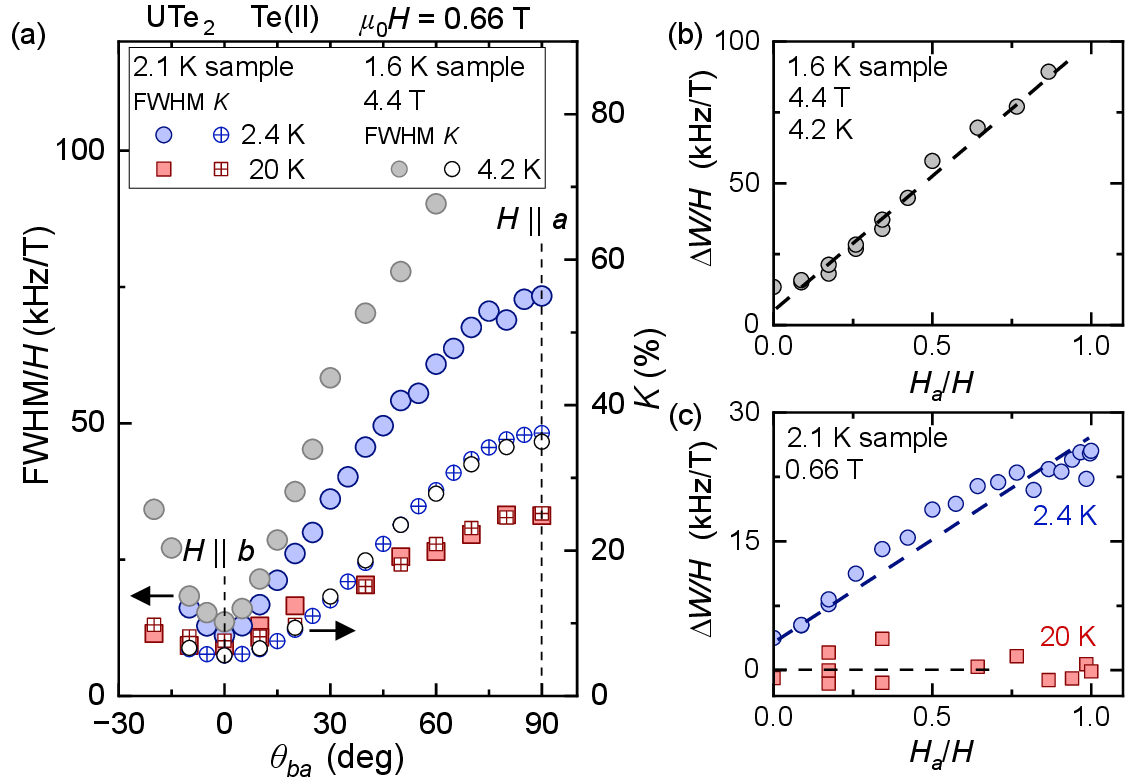}
\caption{
(a) Variation in linewidth of the NMR spectrum and Knight shift at the Te(II) site against the magnetic field angle in the $ab$-plane measured at 2.4 K and 20 K.
The applied magnetic field is 0.66 T.
The broken lines indicate $\theta_{ba} = 0^\circ (H \parallel b)$ and $90^\circ (H \parallel a)$.
We also plotted the result at 4 K and 4.4 T in the 1.6~K samples with gray color for comparison.
The $a$-axis projected field dependence of the anomalous broadening in FWHM $\Delta W$ divided by $H$ in the 1.6 K sample (b) and the 2.1 K sample (c).
The broken lines are guides for the eye.
}
\label{Fig.3}
\end{figure}

\begin{figure*}[!tb]
\centering
\includegraphics[width=18cm,clip]{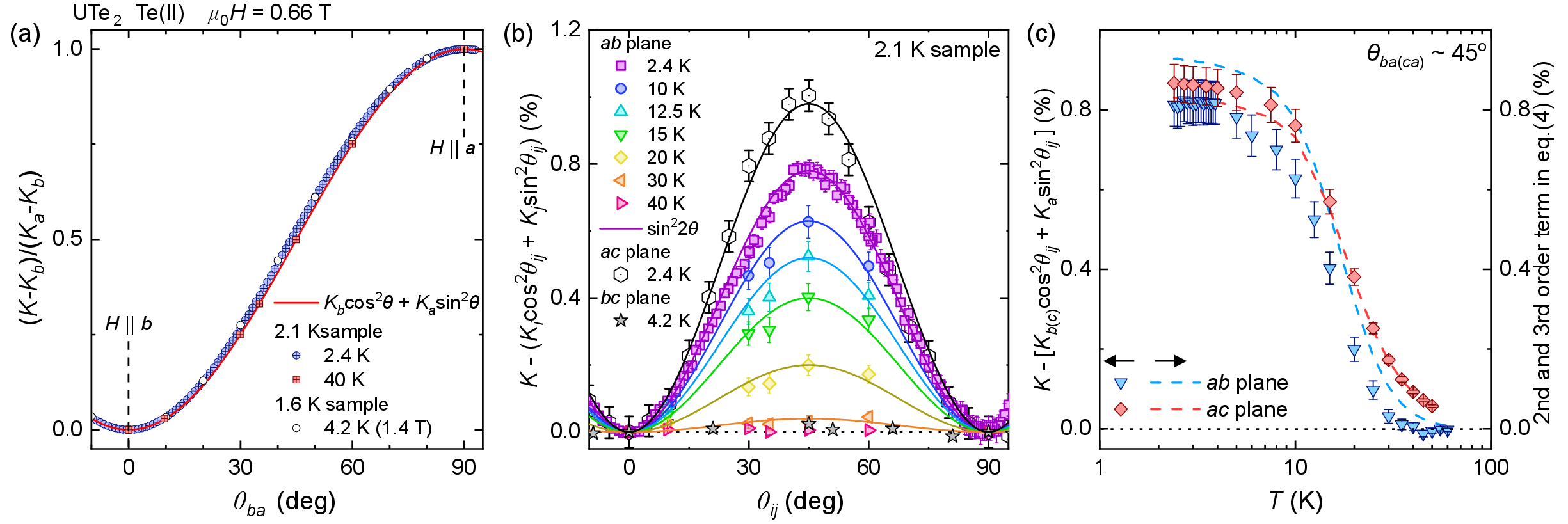}
\caption{
(a) Magnetic field angle dependence of Knight shift normalized by $K_a$ and $K_b$ at 2.4 K and 40 K for the 2.1 K sample and 4.2 K for the 1.6 K sample.
The applied magnetic field is 0.66 T and 1.4 T in the 2.1 K and the 1.6 K sample, respectively.
The broken lines indicate $\theta = 0^\circ (H \parallel b)$ and $90^\circ (H \parallel a)$.
The solid curve indicates theoretical equation for conventional materials ($K_{\theta}^{\rm normal} = K_b \cos^2 \theta + K_a \sin^2 \theta$).
(b) Magnetic field angle dependence of anomalous part of Knight shift at various temperatures in the 2.1 K sample.
We also plot the data rotated in the $ac$ and $bc$ planes.
The solid curves indicate the $\sin^22\theta_{ij}$ curve.
(c) Temperature dependence of anomalous part of Knight shift at $\theta_{ba(ca)} = 45^{\circ}$ and the second- and third-order term in eq.~\eqref{eq:rotation2}.
The dotted lines indicate $K$ = 0\%.
}
\label{Fig.3-2}
\end{figure*}

The most significant contrast between the 1.6 K and 2.1 K samples lies in the linewidth of the NMR spectrum along the $a$ axis.
Previous investigations revealed a substantial increase in linewidth, relative to the Knight-shift value, below 30 K for the 1.6 K sample\cite{Y.Tokunaga_JPSJ_2022,H.Fujibayashi_JPSJ_2023}.
The linewidth of the NMR spectrum is a physical quantity that reflects the distribution of magnetic susceptibility, and in the paramagnetic state, the magnetic susceptibility is proportional to the Knight shift.
Consequently, the substantial increase in linewidth compared to the Knight shift suggests the presence of the distribution of staggered magnetic susceptibility.
As shown in Fig.~\ref{Fig.2}(a), in the 2.1 K sample, an increase in FWHM at lower temperatures is also observed compared to the Knight shift, although this difference is smaller than that observed in the 1.6 K sample\cite{Y.Tokunaga_JPSJ_2022}. 
Previous studies have proposed that this increase in FWHM is caused by the breaking of the magnetic correlations due to the tiny amount of U deficiency.
In the 2.1 K sample, the reduction in the U deficiency due to improved sample quality seems to suppress the increase in FWHM, providing a consistent explanation.
It is noted that, for $H \parallel b$ and $c$, such an effect was smaller, although FWHM is not proportional to $K$ at low temperatures, as shown in Figs.~\ref{Fig.2}(b) and \ref{Fig.2}(c).
In particular, the FWHM is larger than the Knight shift at low temperatures for $H \parallel c$.
This is due to the overlap between the Te(I)- and Te(II)-NMR spectra, which makes the evaluation more difficult.

The anomalous increase in linewidth was observed in magnetic field angle dependence as well.
Figure \ref{Fig.3}(a) illustrates the variation in FWHM of the NMR spectrum and Knight shift at the Te(II) site against the magnetic field angle in the $ab$-plane measured at 2.4 K and 20 K.
We also plotted the result at 4 K in the 1.6 K sample for comparison.
At 20~K, the observed angular dependence of linewidth around the $b$ axis is smooth and scaled with the Knight shift, which is expected in the normal state.
In contrast, at 2.4 K, a kink in linewidth was observed at $H \parallel b$.
This is likely due to the additional linewidth broadening being proportional to the magnetic field along the $a$ axis, as reported in the previous studies\cite{Y.Tokunaga_JPSJ_2022}.
To highlight this additional broadening in FWHM, we plotted $\Delta W = \text{FWHM} - \alpha K$ divided by $H$ as a function of the $a$-axis projected magnetic field $H_a = H\sin \theta_{ba}$, where $\alpha = 0.92$ is the same value between the 1.6 K sample and 2.1 K sample, and $\theta_{ba}$ represents the angle between the magnetic field direction and the $b$ axis. 
As shown in Figs.~\ref{Fig.3} (b) and \ref{Fig.3}(c), $\Delta W/H$ is almost linearly proportional to $H_a$, indicating that the increase is induced by $H_a$.
In the 2.1 K sample, the slope $\Delta W/H_a = 25$~kHz/T is considerably smaller than that in the 1.6 K sample ($\Delta W/H_a = 100$~kHz/T), implying the high quality of the 2.1 K sample.
To clarify the origin of the additional broadening in FWHM, a more detailed investigation into the sample dependence and magnetic field dependence is crucial.
These aspects will be addressed in future work.

\subsection{Unusual angular dependence of Knighit shift due to large magnetic anisotorpy}
From the angular dependence of the Knight shift at low temperatures, we found that the angular rotation from the $b$ axis or $c$ axis to the $a$ axis reflects the large magnetic anisotropy in UTe$_2$.
In general, the Knight shift $K$ is defined as
\begin{align}
K &\equiv \frac{|\bm{H_{\rm ext}}+\bm{H_{\rm hf}}|-|\bm{H_{\rm ext}}|}{|\bm{H_{\rm ext}}|}.
\label{eq.K}
\end{align}
Here, $\bm{H_{\rm ext(hf)}}$ is the applied(hyperfine) magnetic field.
When we assume that the $\bm{H_{\rm hf}}$ is parallel to $\bm{H_{\rm ext}}$, the magnetic field angle dependence of the Knight shift $K_{\theta}$ in the $ij$ plane can be expressed as a projection from each crystal axis component as follows:
\begin{align}
K_{\theta}^{\rm normal} = K_i \cos^2 \theta_{ij} + K_j \sin^2 \theta_{ij}.
\label{eq.angle}
\end{align}
Here, $K_{i(j)}$ is the Knight shift along the $i$($j$) axis, and $\theta_{ij}$ is defined as the angle from the $i$ axis to the $j$ axis.
As shown in Fig.~\ref{Fig.3-2}(a), the angular dependence of the Knight shift in the $ab$ plane of UTe$_2$ agrees well with eq.\eqref{eq.angle} at 40 K, but it deviates at 2.4 K.
To elucidate this difference, we subtracted the normal angle-dependent component, $K_{\theta}^{\rm normal}$, from the experimental data and present the result in Fig.~\ref{Fig.3-2}(b).
The deviation part of the Knight shift increases smoothly with increasing $\theta_{ba}$, peaks at $\theta_{ba} = 45^\circ$, and then decreases toward the $a$ axis, following a $\sin^2 2\theta_{ba}$.
As shown in Figs.~\ref{Fig.3-2}(a) and~\ref{Fig.3-2}(b), such an angular dependence is also observed in the $ac$ plane, at the Te(I) site (not shown), and in the 1.6 K sample, but not in the $bc$ plane.
Although the higher-order angle-dependent term is sometimes discussed as the emergence of nonlinear magnetic susceptibility\cite{R.Okazaki_Science_2011,H.Ikeda_NP_2012,A.Inda_JPSJ_2023,S.Hayami_JPSJ_2024}, we consider that this anomaly on UTe$_2$ originates from the directional difference between $\bm{H_{\rm hf}}$ and $\bm{H_{\rm ext}}$ caused by the large anisotropy of the Knight shift.
Using eq.\eqref{eq.K} and $H_{{\rm hf},i} = K_i H_{{\rm ext},i}$\cite{Y.Tokunaga_JPSJ_2019,H.Fujibayashi_JPSJ_2023}, $K_{\theta}$ in the $ab$ plane can be written as,
\begin{align}
K_{\theta} = \sqrt{(1+K_b)^2 \cos^2\theta_{ba} + (1+K_a)^2 \sin^2\theta_{ba}}-1.
\end{align}
Expanding in a Maclaurin series assuming $K_a$ and $K_b$ are small, we get that
\begin{align}
K_{\theta} \sim &K_b \cos^2\theta_{ba} + K_a \sin^2\theta_{ba}\notag \\
+ &\frac{1}{8} (K_a - K_b)^2 \sin^22\theta_{ba}\notag\\
- &\frac{1}{8} \sin^2 2\theta_{ba} [ K_a^3 \cos^2 \theta_{ba} 
- K_a^2 K_b (3\cos^2 \theta_{ba} - 1) \notag \\ 
&- K_a K_b^2 (3\sin^2 \theta_{ba} - 1) + K_b^3 \sin^2 \theta_{ba} ]
\label{eq:rotation2}
\end{align}
within the third-order expression.
The second-order values of $\frac{1}{8}(K_a - K_b)^2 = 1.2\%$ and $\frac{1}{8}(K_a - K_c)^2 = 1.1\%$ are almost comparable with the amplitude of the $\sin^2 2\theta$ term $\sim 0.8 \%$ and $\sim 0.9 \%$, respectively.
A slight difference may come from the misalignment of the angle and/or the higher-order term contributions.
On the other hand, the value of $\frac{1}{8}(K_c - K_b)^2 = 0.01\%$ is negligible.
Figure~\ref{Fig.3-2}(c) shows the temperature dependence of the anomalous part of the Knight shift at $\theta \sim 45^\circ$ together with the second- and third-order term in eq.~\eqref{eq:rotation2}.
These components develop below 40 K and saturate below 4~K, which are similar to each other.
The magnetic anisotropy of UTe$_2$ is exceptionally large, making conventional approximations inapplicable and suggesting that the higher-order angular dependence is observed.

\begin{figure*}[!tb]
\centering
\includegraphics[width=15cm,clip]{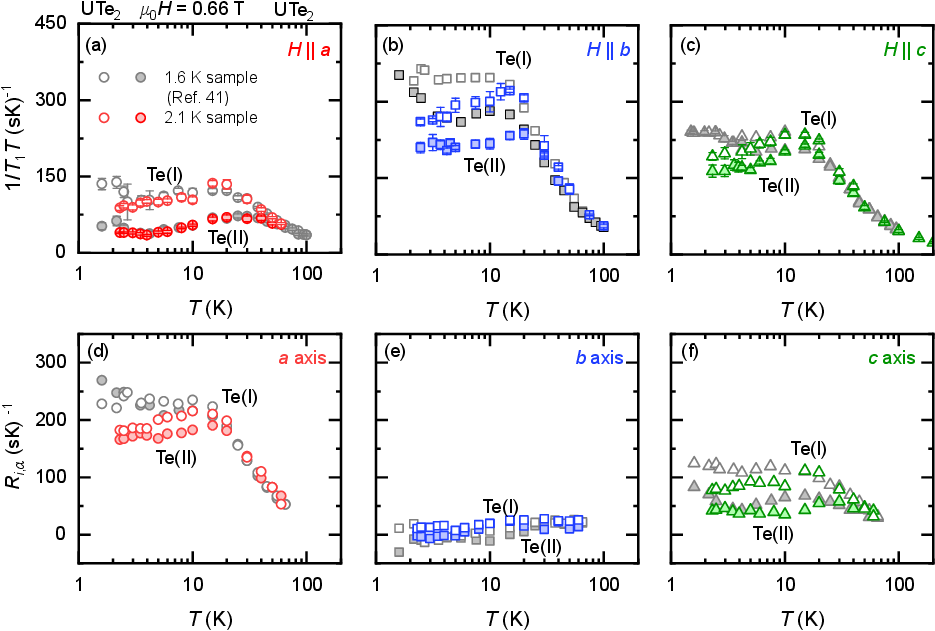}
\caption{
Temperature dependence of $1/T_1T$ at Te(I) and Te(II) for (a)$H \parallel a$, (b)$H \parallel b$, and (c)$H \parallel c$ in the 2.1 K sample of UTe$_2$.
We also plotted the results in the 1.6~K samples with gray color for comparison.
Temperature dependence of (d) $R_{i,a}$, (e) $R_{i,b}$, and (f) $R_{i,c}$ ($i$ = I and II) evaluated from the data shown in the upper panel.
}
\label{Fig.4}
\end{figure*}

\begin{figure}[!tb]
\centering
\includegraphics[width=7.5cm,clip]{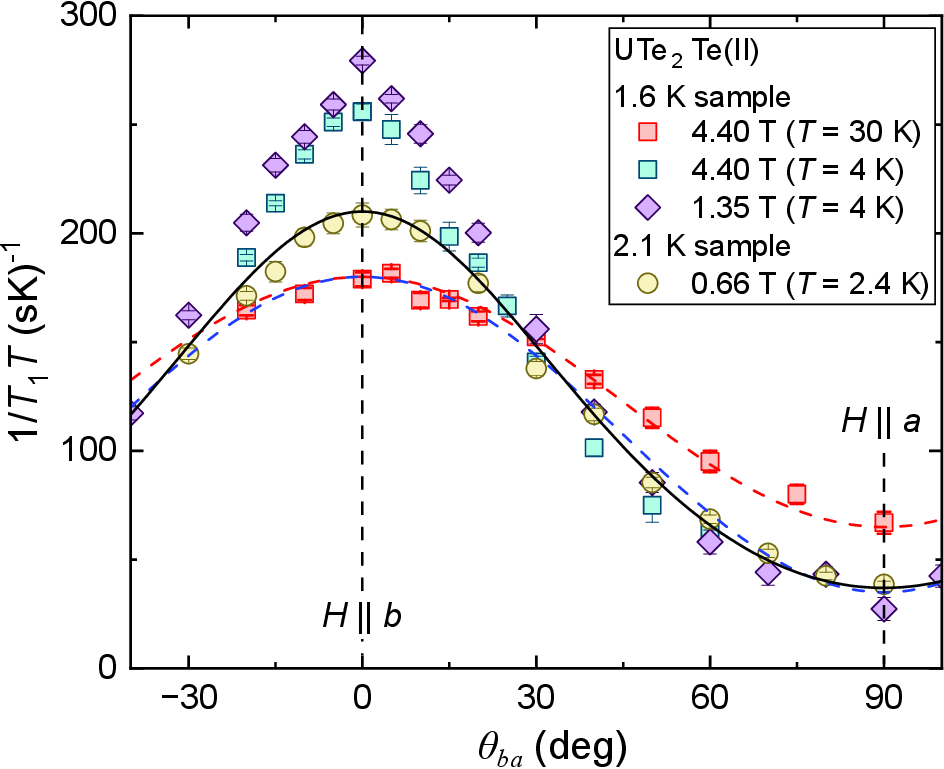}
\caption{
Variation in $1/T_1T$ at the Te(II) site against the magnetic field angle in the $ab$-plane for various conditions.
The broken lines indicate $\theta_{ba} = 0^\circ (H \parallel b)$ and $90^\circ (H \parallel a)$.
The broken~(solid) curves are theoretical formulas expressed as Eq.\eqref{eq:angT1}~[\eqref{eq:angT1-2}].
}
\label{Fig.5}
\end{figure}

\subsection{Magnetic fluctuations}
There are also differences in magnetic fluctuations observed in the 1.6 K and 2.1 K samples.
Figures~\ref{Fig.4} (a)-(c) depict the temperature dependence of the nuclear spin-lattice relaxation rate $1/T_1T$ for each crystallographic axis ($a, b, c$) along with the data in the 1.6 K sample\cite{H.Fujibayashi_JPSJ_2023} for comparison.
In the 2.1 K sample, the sharper NMR linewidth allows us to measure the site dependence of $1/T_1$ for $H \parallel c$ at low temperatures.
While the high-temperature behavior is consistent across the samples, there are clear distinctions in the low-temperature region.
For the $b$ axis, $1/T_1T$ in the 1.6 K sample is almost constant below 10 K\cite{H.Fujibayashi_JPSJ_2023,D.Ambika_PRB_2022,K.Kinjo_PRB_2022}, whereas that in the 2.1 K sample, especially at the Te(I) site, exhibits a broad maximum at around 15~K with relatively smaller values. 
This temperature might be related to a broad Schottky-like anomaly observed in the various thermodynamic, transport probes\cite{K.Willa_PRB_2021,G.Knebel_PRB_2024,Y.Tokiwa_PRB_2024} and the peak of the nuclear spin-spin relaxation rate $1/T_2$\cite{Y.Tokunaga_JPSJ_2022}.
Additionally, the increase observed below 5 K in the 1.6 K sample is absent in the 2.1 K sample.

In general, $1/T_1T$ detects magnetic fluctuations perpendicular to the magnetic field expressed as,
\begin{align}
R_{i,\alpha} = \sum_{\bm{q},\xi}A_{i}^{\alpha\xi}(\bm{q})^2\frac{\chi_{\xi}''(\bm{q},\omega_{\rm N})}{\omega_{\rm N}},
\end{align}
where $\omega_N$ is the NMR frequency, $\chi_{\xi}''(\bm{q},\omega_{\rm N})$ is the imaginary part of the dynamics susceptibility along $\xi = {a, b, \text{and}~c}$, and $A_{i}^{\alpha\xi}(\bm{q})$ is a $\bm{q}$-dependent hyperfine coupling tensor at the Te($i$) ($i$ = I, and II) site.
Using this equation, magnetic fluctuations along each axis ($a, b, c$) can be extracted from each $1/T_1T$ data, as shown in Figs.~\ref{Fig.4} (d)-(f) with results for the 1.6 K sample.
The $b$-axis spin fluctuations are small, resulting in no observable sample dependence.
On the other hand, fluctuations along the $a$ and $c$ axes are suppressed in the 2.1 K sample compared to the 1.6K sample.
This suggests that the enhancement in the $a$- and $c$-axes magnetic fluctuations below 4~K is associated with the U deficiency and that the high quality of the 2.1 K sample led to the suppression of these fluctuations.
The suppression of low-temperature fluctuations in high-quality samples was also observed in recent muon spin relaxation measurements\cite{S.Sundar_PRB_2019,N.Azari_PRL_2023}.

As the site dependence of $1/T_1$ under the $c$-axis magnetic field can be measured in the 2.1 K sample, we reveal that $a$-axis fluctuations also exhibit site dependence.
The temperature dependence of $R_{i,\alpha}$ roughly corresponds to that of the Knight shift\cite{H.Fujibayashi_JPSJ_2023}, suggesting the predominance of ferromagnetic ($q = 0$) correlations.
On the other hand, for the $a$-axis, the static magnetic susceptibility at the Te(II) site is larger than that at the Te(I) site\cite{H.Fujibayashi_JPSJ_2023}, while spin fluctuations at the Te(II) site are smaller than those at the Te(I) site.
Furthermore, despite almost the site-independent static magnetic susceptibility along the $c$ axis, $c$-axis spin fluctuations are more pronounced at the Te(I) site.
It is considered that the difference between the static susceptibility and the spin fluctuations originates from the difference in the $q$-dependent hyperfine coupling tensor $A_{i}^{\alpha\xi}(\bm{q})$.
While the static susceptibility is related to the $q = 0$ component, the spin fluctuation detects all $q$ components.
Therefore, these experimental results suggest the coexistence of a ferromagnetic and an antiferromagnetic correlation with $q \neq 0$ in UTe$_2$, which is consistent with the antiferromagnetic fluctuations with $Q$ = (0, 0.57, 0) observed in the neutron measurements\cite{W.Knafo_PRB_2021}.

The low-temperature magnetic fluctuations in the 1.6~K sample are also reflected in the magnetic field angle dependence of $1/T_1T$, as shown in Fig.~\ref{Fig.5}.
The angle dependence of $1/T_1T(\theta)$ typically exhibits similar angle dependence to the Knight shift in the normal state.
In the 1.6 K sample, the experimental data at 30~K can be fitted well with a theoretical formula described as,
\begin{align}
\frac{1}{T_1T}(\theta_{ba}) = \left(\frac{1}{T_1T}\right)_{H\parallel b}\cos^2 \theta_{ba} + \left(\frac{1}{T_1T}\right)_{H\parallel a}\sin^2 \theta_{ba}.
\label{eq:angT1}
\end{align}
However, at 4 K, a kink behavior around the $b$ axis was observed.
This suggests the enhancement of magnetic excitations specifically around the $b$ axis.
Similarly to the additional increase in linewidth, the kink behavior in $1/T_1T$ is also attributed to the presence of the U deficiency.
However, as shown in Fig.~\ref{Fig.5}, the anomaly in $1/T_1T$ shows the same angle dependence, regardless of the magnetic field magnitude, indicating that it is related not to the magnitude of $H_a$ but to the angles, unlike the case of the additional increase in linewidth.
This result indicates that the enhancement of magnetic excitation around the $b$-axis is neither induced nor suppressed by $H_a$, in contrast to the behavior of UCoGe, where ferromagnetic fluctuations along the magnetic easy axis are suddenly suppressed with increasing the $c$-axis magnetic field\cite{T.Hattori_PRL_2012,K.Ishida_PRB_2021}.

In the case of the 2.1 K sample, such kink behavior in $1/T_1T$ was not observed even at 2.4 K.
Here, the angular dependence of $1/T_1T$ at 2.4 K in the 2.1 K sample is thought to arise from the deviation between the directions of $\bm{H_{\rm ext}}$ and the effective magnetic field $\bm{H_{\rm eff}} = \bm{H_{\rm ext} +\bm{H_{\rm hf}}}$ due to the anisotropy of the Knight shift, similarly to the case of the angular dependence of the Knight shift. 
As shown in the solid curve of Fig.~\ref{Fig.5}, the angular dependence of $1/T_1T$ in the $ab$ plane can be well explained by the following equation:
\begin{align}
\frac{1}{T_1T}(\theta_{ba}) = \left(\frac{1}{T_1T}\right)_{H\parallel b}\cos^2 \theta' + \left(\frac{1}{T_1T}\right)_{H\parallel a}\sin^2 \theta'.
\label{eq:angT1-2}
\end{align}
Here, $\theta' = \tan^{-1}\left(\frac{1+K_a}{1+K_b}\tan \theta_{ba}\right)$ is the angle between $\bm{H_{\rm eff}}$ and the $b$ axis.

\subsection{Summary of low-temperature magnetic properties}
Finally, we summarize how the magnetic properties are different between the 1.6 K and 2.1 K samples.
The static magnetic susceptibility and high-temperature magnetic fluctuations are relatively unaffected by the enhancement of sample quality.
However, the distribution of $a$-axis magnetic susceptibility and 1/$T_1T$ in $H \parallel b$ exhibit significant differences between the samples.
This supports the presence of slow magnetic fluctuations induced by the U deficiency, as previously proposed in the literature\cite{Y.Tokunaga_JPSJ_2022}.
Such slow fluctuations might be related to the residual electronic term in the specific measurement.
These results indicate that in UTe$_2$, the quality of the sample is quite important to reveal the effective interaction for superconductivity.
The improvement in sample quality also extends the region of the high-field SC phase\cite{Z.Wu_PNAS_2024,D.Aoki_JPSJ_2024b}, suggesting that measurements of magnetic properties on the 2.1 K sample will provide insight into the origin of the high-field SC phase.
Moreover, we found that the angular rotation in the $ab$ and $ac$ planes does not exhibit the ordinary angular dependence and that a higher-order $\sin^22\theta$ term appears below 40 K, which is related to huge magnetic anisotropy in UTe$_2$. 

\section{Conclusion}
In conclusion, we performed $^{125}$Te-NMR measurements in the ultra-clean 2.1~K sample of UTe$_2$ and compared the results with those in the 1.6~K sample to investigate the intrinsic magnetic properties of UTe$_2$.
The enhancement of the linewidth of the NMR spectrum in the $a$-axis magnetic field and the low-temperature magnetic fluctuations observed in the 1.6~K sample are suppressed in the 2.1 K sample, suggesting that a tiny amount of U deficiency induces such anomalous magnetic properties.
Our findings highlight the importance of high-quality samples for investigating the essential magnetic and SC properties of UTe$_2$.

\section*{acknowledgments}
We acknowledge S. Yonezawa and Y. Yanase for fruitful discussion. 
This work was supported by Grants-in-Aid for Scientific Research (KAKENHI Grant No. JP20KK0061, No. JP20H00130, No. JP21K18600, No. JP22H04933, No. JP22H01168, No. JP23H01124, No. JP23K22439 and No. JP23K25821) from the Japan Society for the Promotion of Science, by JST SPRING(Grant No. JPMJSP2110) from the Japan Science and Technology Agency, by research support funding from the Kyoto University Foundation, by ISHIZUE 2024 of Kyoto University Research Development Program, by Murata Science and Education Foundation, and by the JGC-S Scholarship Foundation.
Liquid helium is supplied by the Low Temperature and Materials Sciences Division, Agency for Health, Safety and Environment, Kyoto University.

H.M. and S.K. equally contributed to this work.

%

\end{document}